\documentclass[aps, showpacs,twocolumn,superscriptaddress]{revtex4}
\usepackage{amssymb}
\usepackage{natbib}
\usepackage{amsmath}
\usepackage{amsfonts}
\usepackage{graphicx}
\usepackage{mathrsfs}
\usepackage{dcolumn}
\usepackage{bm}
\usepackage{color}
\usepackage{cancel}
\usepackage{mathbbol}
\usepackage{epsfig}
\usepackage{units}
\usepackage{esint}
\usepackage{soul}

\definecolor{rot}{rgb}{0.75,0.05,0.25}
\definecolor{hellgrau}{gray}{0.5}
\definecolor{blau}{rgb}{0,0,0.7}

\newcommand{\sm}{\scriptscriptstyle}
\newcommand{\avl}{\big\langle}
\newcommand{\avr}{\big\rangle}

\newcommand{\be}{\beta}

\newcommand{\ch}{\mathrm{C}}

\newcommand{\ev}{U}

\newcommand{\id}{\mathbb{I}}

\newcommand{\drm}{d}
\newcommand{\e}{\mathop{\mathrm{e}}\nolimits}
\newcommand{\iu}{i\,}
\newcommand{\sd}{J}

\newcommand{\w}{\omega}

\newcommand{\tme}{s}
\def\Tr{\mbox{Tr}}

\def\Tr{\mbox{Tr}}

\def\Tr{\mbox{Tr}}

\begin{document}

\title{Geometric quantum pumping in the presence of dissipation}

\author{Juzar Thingna}
\affiliation{University of Augsburg, Institute of Physics, Universit\"atsstrasse 1, D-86135 Augsburg, Germany}
\affiliation{Nanosystems Initiative Munich, Schellingstrasse 4, D-80799 M\"{u}nchen, Germany}

\author{Peter H\"anggi}
\affiliation{University of Augsburg, Institute of Physics, Universit\"atsstrasse 1, D-86135 Augsburg, Germany}
\affiliation{Nanosystems Initiative Munich, Schellingstrasse 4, D-80799 M\"{u}nchen, Germany}
\affiliation{Physics Department, Blk S12, National University of Singapore, 2 Science Drive 3, 117551 Singapore}

\author{Rosario Fazio}
\affiliation{NEST, Scuola Normale Superiore \& Istituto Nanoscienze-CNR, I-56126 Pisa, Italy}
\affiliation{Centre for Quantum Technologies, National University of Singapore, 3 Science Drive 2, 117543 Singapore }

\author{Michele Campisi}
\affiliation{University of Augsburg, Institute of Physics, Universit\"atsstrasse 1, D-86135 Augsburg, Germany}
\affiliation{Nanosystems Initiative Munich, Schellingstrasse 4, D-80799 M\"{u}nchen, Germany}
\affiliation{NEST, Scuola Normale Superiore \& Istituto Nanoscienze-CNR, I-56126 Pisa, Italy}

\date{\today }

\begin{abstract}

The charge transported when a quantum pump is adiabatically driven by time-dependent external forces in presence of dissipation is given by the line integral of a pumping field $\bm{F}$. We give a general expression of $\bm{F}$ in terms of quantum correlation functions evaluated at fixed external forces. Hence, an advantage of our method is that it transforms the original time-dependent problem into an autonomous one. Yet another advantage is that the curl of $\bm{F}$ gives immediate visual information about the geometric structures governing dissipative quantum pumping. This can be used in a wide range of experimental cases, including electron pumps based on quantum dots and Cooper-pair pumps based on superconducting devices. Applied to a Cooper-pair sluice, we find an intriguing dissipation-induced enhancement of charge pumping, reversals of current, and emergence of asymmetries. This geometric method thus enables one to unveil a plethora of beneficial, dissipation-assisted operation protocols.
\end{abstract}


\maketitle
\section{Introduction}
Ever since the discovery of the geometrical phase accompanying adiabatic
driving in quantum systems \cite{Berry84PRSLA392}, the role that geometric quantities play in
many physical  phenomena has been in the focus of intense research 
\cite{Resta94RMP66,Xiao10RMP10,Sinitsyn09JPA42,Wilczek89Book}. By geometric quantities we mean here
quantities which are determined solely by the geometry of the path drawn by the
changing driving parameters. The Berry phase is one such quantity: As a quantum system is
adiabatically transported along a closed cycle in the space of the driving parameters, its wave
function accumulates a phase $\Theta_G$ which depends only on the geometry of the cycle. In
particular, $\Theta_G$ is the line integral of a vector field (the Berry connection)
over the closed path in the parameter space.
Geometric quantities are indeed common in other branches
of physics besides quantum mechanics, and all can be expressed as the line integral of some
vector field. The most prominent example is the work output $W=\oint dV P$
per cycle of a thermodynamic engine \cite{Sinitsyn09JPA42,Hannay06AJP74}.
This is perhaps the simplest example of a geometric pump, namely, a
system that adiabatically converts an ac driving into a dc current (not to be confused with rectification). Like the thermodynamic engine,
any geometric pump is fully characterized by a vector field $\bm F$, which we shall call the
pumping field.

Adiabatic pumps are currently in the limelight of topical experimental and theoretical research.
Stochastic pumps~\cite{Sinitsyn09JPA42}, whose mechanisms underlie, e.g., the functioning of Brownian
motors~\cite{Hanggi09RMP81}, or heat pumps~\cite{Ren10PRL104} are important examples.
Quantum charge pumps~\cite{Thouless83PRB27,Brouwer98PRB58,Zhou99PRL82}, based on the
adiabatic manipulation of coherent devices, are another exciting avenue of research  of this kind,
also in view of their application to metrology~\cite{Pekola13RMP85}.  Since the pioneering paper
by Thouless~\cite{Thouless83PRB27} many  aspects of adiabatic pumping have been elucidated.
An incomplete list includes the scattering theory of (charge, spin, heat) pumping~\cite{Thouless83PRB27,
Brouwer98PRB58,Aleiner98PRL81,Entin-Wohlman02PRB65,Moskalets02PRB66},
its extension to include electron-electron interaction~\cite{Citro03PRB68,Splettstoesser05PRL95,Sela06PRL96,Fioretto08PRL100},
the theory of Cooper-pair pumping in superconducting nanocircuits~\cite{Pekola01PRL64,Governale05PRL95,Mottotten06PRB73},
and topological pumping~\cite{Thouless83PRB27,Onoda06PRL97}.  Along with this intense theoretical activity, a number of important experiments
have been successfully performed~\cite{Pothier92EPL17,Mottonen08PRL100,Pekola08NATPHYS4,Giazotto11NATPHYS7}.
In all these cases, dissipation plays an unavoidable, possibly constructive role, whose features are yet to be fully understood.
This motivated a renewed interest  in studying the combined effects of noise and driving \cite{Grifoni98PR304} in the context of adiabatic
quantum transport \cite{Brandes02PRB66,Salmilehto10PRA82,Solinas10PRB82,Pekola10PRL105,Russomanno11PRB83,
Kamleitner11PRB84,Pellegrini11PRL107,Wollfarth13PRB87}.

All those prior attempts attacked the problem by solving the reduced dynamics of the slowly driven
open quantum system within some approximation scheme appropriate to each specific physical case.
This gives the reduced density matrix $\rho_t$, which is used to calculate the instantaneous current
$\mathcal I= \Tr I \rho_t$, and by time integration the total pumped charge, out of which one has to single out the geometric contribution.
Here we pursue instead {\it a geometric approach} to calculate the pumping field
$\bm F$ giving the geometrically pumped charge directly.
Our approach is based on the salient  observation that,  independent of the specific physical scenario, $\bm F$ is in fact the
vector of linear response coefficients in the adiabatic expansion of the current [see Eq. (\ref{eq:Taylor}) below].
To the best of our knowledge, this  result was never exploited before in the context of dissipative quantum pumping. 
It brings about two main advantages: (i) When applied to an open quantum system, it leads to an exact expression of $\bm{F}$ in terms of
equilibrium quantum correlation functions, which are calculated at frozen driving parameters.
That is, our scheme makes evident that solving the {\it reduced} dynamics of an undriven system suffices.
(ii) Since $\bm F$ characterizes the geometric
pumping fully, once one knows it, calculating the charge pumped along any cycle is as simple as
doing a line  integral. Besides, the curl $\bm G= \bm \nabla \times \bm F$ provides immediate
visual information about the geometric features associated to dissipative pumping. These
unveil the possibility of many previously undetected dissipation-assisted operation protocols.
See our Cooper-pair sluice
example below.

Our expression of the pumping field [see Eq. (\ref{eq:F})] can be used in a wide range of cases of experimental interest, ranging
from electron pumps based on quantum dots to Cooper-pair pumps in superconducting devices. We illustrate
the method in the latter case. To this end we pursue here the derivation of a specific equation of motion (EOM) for the
calculation of  equilibrium quantum correlation functions at fixed driving parameters under the sole  assumption
of weak coupling to a bosonic bath. We emphasize that the EOM neither rests on a Markov nor
a rotating wave approximation.


\section{Theory}
Consider a generic geometric pump, namely, a physical device that can be externally manipulated by several control parameters $\bm B= (B_x,B_y,B_z, \dots)$ and supports the flow of a current $\mathcal{I}$.
We are interested in the ``charge'' $q= \int_0^\mathcal{T} dt \mathcal{I}_t$ that is transported by the device as the parameters
draw a closed path $\mathcal C$ in the parameter space. The symbol  $\mathcal T$ denotes the time duration of the cycle. In general, the transported charge has a geometric contribution. The key to singling out the geometric component of the transported charge is to perform an ``adiabatic expansion'' of the current at any generic time $t$, namely, a Taylor expansion of the current $\mathcal{I}$ in terms of the rate of change $\dot{\bm B}$ of the parameters:
\begin{equation}
\mathcal{I} = \mathcal{I}_0 + {\bm F} \cdot \dot{\bm B}+ \sum_{i,j} L_{ij} \dot{B}_i \dot{B}_j + \dots ,
\end{equation}
where the coefficients $F_i, L_{ij}, \dots$ are functions of the value $\bm B$ taken by the parameters at time
$t$. Accordingly, the transported charge is given by
\begin{equation}
q = \int_0^\mathcal{T} \mathcal{I}_0 dt + \int_0^\mathcal{T}dt {\bm F} \cdot \dot{\bm B} + \int_0^\mathcal{T} dt \sum_{i,j} L_{ij} \dot{B}_i \dot{B}_j  \, + \dots .
\label{eq:Taylor}
\end{equation}
The zeroth-order term is what is customarily referred to as the dynamical charge. It is due to the fact that
charge could possibly flow even at fixed parameters \cite{footnote}.
Note that the dynamical charge depends very strongly on the duration of the cycle: the same cycle operated at half the speed would result in twice the dynamic charge. The first-order term, in contrast, is geometric, because
$
\int_0^\mathcal{T} d{t}\bm F(\bm B_{t}) \cdot \dot{\bm B}_{t} = \oint \bm F(\bm B) \cdot d
\bm B $ depends only on the geometry of the path. On the contrary, the higher-order
terms are not geometric, because the change of variable $\dot{\bm B} dt = d\bm B$ would not suffice to remove the explicit $\dot{\bm B}$ dependence of their integrands. Thus the full geometric contribution to the transported charge $q_G$ is exactly and solely given by
\begin{align}
q_G = \oint_\mathcal{C} \bm F(\bm B) \cdot d \bm B = \oiint \bm G(\bm B) \cdot d \bm \Sigma \;,
\label{eq:geocharge}
\end{align}
with $\bm{F}$ the vector of adiabatic linear response coefficients, and $\bm G= \bm \nabla \times \bm F$ its curl. (The double integral is a surface integral over any surface having $\mathcal C$ as its contour, i.e., the Stokes theorem.)

\section{The Cooper pair sluice}
\begin{figure}[]
		\includegraphics[width=\linewidth]{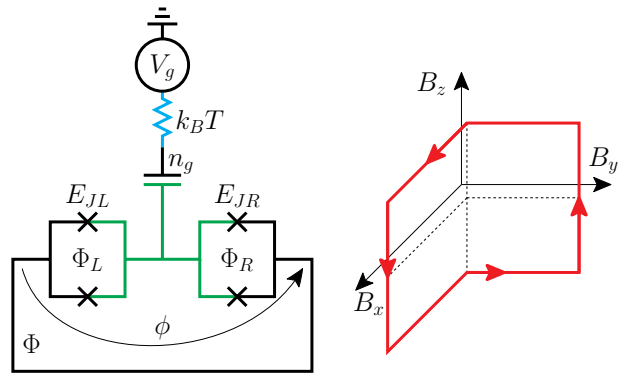}
		\caption{(Color online) Left panel: Schematics of the Cooper-pair sluice.  Two SQUIDS
of tunable Josephson couplings $E_{JL}(\Phi_L)$, $E_{JL}(\Phi_R)$ are separated by a
superconducting island (green) whose polarization charge $n_g$ is externally controlled by the
gate voltage $V_g$. The resistor (blue) represents environmental gate noise of thermal energy
$k_B T$. The threading magnetic flux $\Phi$ fixes the overall phase difference $\phi$ across the
sluice. Right panel: Typical driving path used in experiments \cite{Mottonen08PRL100}.}
		\label{fig:sluice}
\end{figure}
As an application of timely interest, we consider the Cooper-pair sluice   \cite{Mottonen08PRL100} sketched in
Fig. \ref{fig:sluice}.
The sluice consists of two superconducting quantum interference devices (SQUIDS) separated by a superconducting island. The system is phase biased, with the
phase difference $\phi$.
The two SQUIDS with respective Josephson  couplings $E_{JL}$
and $E_{JR}$ can be independently manipulated by controlling the magnetic flux threading each of
them, $E_{JL}= E_{JL}(\Phi_L)$ and $E_{JR}=E_{JR}(\Phi_R)$. The island is further capacitively
coupled to a gate electrode controlling its polarization charge in units of Cooper pairs
$n_g=C_gV_g/2e$, where $C_g$ is the gate capacitance, $e<0$ is the electron's charge, and
$V_g$ is the applied gate voltage. The  three driving parameters in this case are
$B_x=E_{JL}$ , $B_y=E_{JR}$, and $B_z= E_C(1-2n_g)$, with $E_C$ the charging energy of the island.
We assume the sluice is operated in the regime where the charging energy $E_C$ is
much larger than $E_{JL}$ and $E_{JR}$. In this regime the sluice can be conveniently modeled
as a tunable two-level system. In the basis of charge states $\{|0\rangle, |1\rangle\}$, the Hamiltonian
reads
$H(\bm B) = -\bm B \cdot \bm S $,
with
$S_x =[\sigma_{x}\cos (\nicefrac{\phi}{2})+\sigma_{y}\sin(\nicefrac{\phi}{2})]/2$,
$S_y=[ \sigma_{x}\cos(\nicefrac{\phi}{2})-\sigma_{y}\sin(\nicefrac{\phi}{2})]/2$, and
$S_z= \sigma_{z} /2$ ($\sigma_x,\sigma_y,\sigma_z$ are the Pauli matrices). The charge flowing
through the sluice is associated to the current operator \cite{Aunola03PRB68}
\begin{align}
I =  \frac{2e}{\hbar}\partial_\phi H \, .
\label{eq:I}
\end{align}
The sluice is subject to thermal noise at temperature $T$ coming from the voltage gate.

Depending on the time scale $\mathcal{T}$, there are two distinct adiabatic regimes:
the coherent regime and the dissipative regime \cite{Kamleitner11PRB84}.
The first is relevant when the driving time $\mathcal{T}$
is very short compared to the thermal relaxation and decoherence times but long compared to the transition times $
\Omega_{mn}^{-1}(\bm B)$. Accordingly, one performs an adiabatic expansion around the eigenstate $|n(\bm B)
\rangle$ (as in \cite{Berry93PRSLA442}), and the pumping field emerges as
$
\bm F_n(\bm B) = {2}{ \hbar}^{-1} \,  \text{Im} \sum_{k(\neq n)}{\bm S_{nk}(\bm B) I_{kn}(\bm
B)}{\Omega_{kn}^{-2}(\bm B)}.
$
For the Cooper pair sluice working close to the ground state, we obtain, for its curl, the following analytical
result:
\begin{align}
\bm G_0(\bm B) =-e\frac{ \cos \phi \left(\bm B^2-B_x B_y \cos \phi \right)+3 B_x B_y}
{\left(\bm B^2 +2 B_x B_y \cos \phi\right)^{5/2}} \bm B\;.
\label{eq:G0}
\end{align}
Notice that $\bm G_0=-2e\partial_\phi \bm B_{B}$, where $\bm B_B$ is the ground-state Berry
curvature \cite{Aunola03PRB68}.

In the case when the driving time $\mathcal{T}$ is very short compared to the thermal-relaxation time, instead, the reference state around which the expansion is performed is the instantaneous thermal state $\rho_{\bm B}^\text{eq}=e^{-\beta \mathcal H(\bm B)}/Z(\bm B)$, with $Z(\bm B)$ the partition function and
\begin{equation}
\mathcal{H}({\bm B})= H({\bm B}) +H_E + H_{S E},
\label{totalH}
\end{equation}
the total Hamiltonian, sum of system, environment, and coupling Hamiltonians, respectively.
The adiabatic linear response theory developed in Ref. \cite{Campisi12PRA86} gives then
\begin{align}
\label{eq:F} \bm F(\bm B) &= -\int_{-\infty}^0 ds \int_0^\beta du \Tr \rho_{\bm B}^\text{eq} I_{-i\hbar u} \Delta \bm S_s ,
\end{align}
where $\bm S = -\bm \nabla H$, $\Delta \bm S = \bm S - \Tr \rho_{\bm B}^\text{eq} \bm S$,  and the subscripts of $I$ and $\Delta \bm S$ denote that these operators are considered in the Heisenberg representation generated by the full Hamiltonian  (\ref{totalH}) (with fixed $\bm B$) at the times $-i\hbar u$ and $s$, respectively.
Note that, as anticipated, the correlation functions of Eq. (\ref{eq:F}) are evaluated at fixed $\bm B$. Clearly, their evaluation does not involve the solution of a driven open-system dynamics. The expression (\ref{eq:F}) is exact and approximations enter only at the point of evaluating it. Below we present our original method for its calculation.
We emphasize that Eq. (\ref{eq:F}) cannot be obtained within the common reduced density matrix approach \cite{BreuerPetruccioneBOOK}, because the sole, single-time, reduced density matrix operator does not suffice for the exact evaluation of two-time quantum correlations \cite{Talkner86AP167,Campisi12PRA86}.

\section{Equation of Motion for the Quantum correlation function}
We model the thermal environment of the sluice as a set of harmonic oscillators
\cite{Caldeira83AP149,Hanggi05CHAOS15,Makhlin73RMP01}:
 \begin{align}
H_{E} = \sum_{\alpha = 1}^{\infty} \frac{p_{\alpha}^{2}}{2 m_{\alpha}} + \frac{m_{\alpha}
\omega_{\alpha}^{2}}{2}x_{\alpha}^{2}, \,\, \,
H_{SE} = A \otimes E.
\end{align}
Here $x_\alpha$, $p_\alpha$, $m_\alpha$, $\omega_\alpha$, are the oscillators positions,
momenta, masses, and frequency, respectively, $A$ is a system operator, and $E$ is an
environment operator.
The evaluation of the field $\bm F(\bm B)$ in (7) involves evaluating  the equilibrium quantum correlation function at various, but fixed, parameter values $\bm B$. 
The dependence of $\bm F$ on $\bm B$ comes from the parametric dependence of the total Hamiltonian 
$
\mathcal{H}(\bm B)$
on $\bm B$.
To this end we begin by writing the imaginary-time integrals in Eq. (\ref{eq:F}) as real-time integrals \cite{Hu93AJP61},
$
\bm F(\bm B)=\frac{i}{\hbar} \int_{0}^{\infty} d\tme\,\tme\, \langle[I,\Delta \bm S_{-\tme}]\rangle_{\bm B}^\text{eq}
= \frac{i}{\hbar}\int_0^\infty d\tme\,\,\tme \,\mathrm{Tr}_S
 I (\bm Y_{-\tme}-\bm Y_{-\tme}^{\dagger}) $,
where 
\begin{align}
\bm Y_{-\tme}(\bm B)= \Tr_E \left[\ev_{\tme}(\bm B) \Delta \bm S \rho^\text{eq}_{\bm B}
\,\ev_{\tme}^\dagger(\bm B)\right]\, ,
\end{align}
and $\Tr_{S(E)}$ 
denotes trace over the system (environment) Hilbert space. The operator $\Delta \bm S$ belongs to the system-Hilbert space, while $U_{\tme}(\bm B) = e^{-\frac{i}{\hbar}\mathcal H(\bm B)\tme}$ is the evolution operator with a \emph{fixed}, frozen $\bm B$. For simplicity of notation, we will keep in the following, the parametric dependence on the fixed $\bm B$ implicit.

Our aim is to obtain an equation of motion (EOM) \cite{Grabert80JSP22} for $\bm Y_{-\tme}$.
We first focus on the auxiliary operator in the full-Hilbert space $\bm Y^{{\sm \text{tot}}}_{-s}= \ev_{\tme} \Delta \bm S \rho^\text{eq}_{\bm B}
\,\ev_{\tme}^\dagger
$. Next, using the Kubo identity $\e^{\be(A+B)}=\e^{\be A}\left[ \id + \int_{0}^{\be} \drm\lambda \e^{-\lambda A}B\e^{\lambda(A+B)}\right]$ we expand the evolution operator $\ev_{\tme}$ up to second order in $H_{SE}$ to obtain
\begin{align}
\ev_{\tme}&=\ev^{\sm{0}}_{\tme}\ev^{\sm{I}}_{\tme},\\
\ev^{\sm{0}}_{\tme}&=\e^{-\frac{i}{\hbar} \left(H+H_{E}\right)\tme},\\
\ev^{\sm{I}}_{\tme} &=\id-\frac{\iu}{\hbar} \int_{0}^{\tme}\drm \tme_{\sm{1}} H_{SE}(\tme_{\sm{1}})\nonumber \\
&-\frac{1}{\hbar^{2}}\int_{0}^{\tme}\drm \tme_{\sm{1}} H_{SE}(\tme_{\sm{1}})\int_{0}^{\tme_{\sm{1}}} \drm \tme_{\sm{2}} H_{SE}(\tme_{\sm{2}}),
\end{align}
where $H_{SE}(\tme) = \ev^{\sm{0}\dagger}_{\tme}H_{SE}\ev^{\sm{0}}_{\tme}$ is the free evolution of $H_{SE}$, and $\ev^{\sm{I}}_{\tme}$ is the truncated-evolution operator in the interaction picture. 

Using the above definition of the evolution operator and differentiating $\bm Y^{{\sm \text{tot}}}_{-\tme}$ with respect to $\tme$ we obtain an integro-differential equation,
\begin{align}
\label{eq:YwoI}
\frac{\drm \bm Y^{{\sm \text{tot}}}_{-\tme}}{\drm \tme}&=-\frac{i}{\hbar}[H+H_{E},\bm Y^{{\sm \text{tot}}}_{-\tme}]-\frac{i}{\hbar}[H_{SE},\bm Y^{{\sm \text{tot}}}(-\tme)]\nonumber \\
&-\frac{1}{\hbar^{2}}\int_{0}^{\tme}\drm \tme_{\sm{1}} [H_{SE},[H_{SE}(\tme_{\sm{1}}-\tme),\bm Y^{{\sm \text{tot}}}(-\tme)]],
\end{align} 
involving the operator $\bm Y^{{\sm \text{tot}}}(-\tme)= 
\ev^{\sm{0}}_{\tme} \Delta \bm S \rho^\text{eq}
\,\ev^{\sm{0}\dagger}_{\tme}$. The latter contains information about the system-environment coupling due to the presence of $\rho^\text{eq}$. Hence, we proceed to expand that up to first order. Using the expansion of the equilibrium density matrix
\cite{Talkner09JSM09,Thingna12JCP19}
\begin{align}
\frac{e^{-\beta \mathcal{H}}}{Z} &\simeq \frac{e^{-\beta \left(H + H_{E}\right)}}{Z_{S}Z_{E}}\left[\id - \int_{0}^{\beta}\drm\beta_{{\sm 1}} H_{SE}(-i\hbar\beta_{{\sm 1}})\right],
\end{align}
where $Z_{S} = \Tr_{S}\left(e^{-\beta H}\right)$ and $Z_{E} = \Tr_{E}\left(e^{-\beta H_{E}}\right)$, we obtain
\begin{align}
\bm Y^{{\sm \text{tot}}}(-\tme) & = \tilde{\bm Y}^{{\sm \text{tot}}}(-\tme) - \tilde{\bm Y}^{{\sm \text{tot}}}(-\tme)\int_{0}^{\beta}\drm\beta_{{\sm 1}} H_{SE}(-\tme-i\hbar\beta_{{\sm 1}}),
\end{align}
where $\tilde{\bm Y}^{{\sm \text{tot}}}(-\tme) = \ev^{\sm{0}}_{\tme}\, \Delta\bm S \tilde{\rho}^\text{eq} \, \ev^{\sm{0}\dagger}_{\tme}$ with $\tilde{\rho}^\text{eq} = e^{-\beta H}/Z_{S} \otimes e^{-\beta H_{E}}/Z_{E}$. Using the above expansion in Eq.~(\ref{eq:YwoI}) and keeping terms only up to second order in $H_{SE}$, we find
\begin{align}
\label{eq:YwI}
\frac{\drm \bm Y^{{\sm \text{tot}}}_{-\tme}}{\drm \tme} &=-\frac{i}{\hbar}[H+H_{E},\bm Y^{{\sm \text{tot}}}_{-\tme}]-\frac{i}{\hbar}[H_{SE},\tilde{\bm Y}^{{\sm \text{tot}}}(-\tme)] \nonumber \\
&+ \frac{i}{\hbar}\int_{0}^{\beta}\drm\beta_{{\sm 1}} [H_{SE},\tilde{\bm Y}^{{\sm \text{tot}}}(-\tme)H_{SE}(-\tme-i\hbar\beta_{{\sm 1}})] \nonumber \\
&-\frac{1}{\hbar^{2}}\int_{0}^{\tme}\drm \tme_{\sm{1}} [H_{SE},[H_{SE}(\tme_{\sm{1}}-\tme),\tilde{\bm Y}^{{\sm \text{tot}}}(-\tme)]].
\end{align} 

Tracing over the environment degrees of freedom and using $H_{SE} = \sigma_{z} \otimes \sum c_n x_n = A \otimes E$, we obtain the equation of motion for the reduced operator $\bm Y_{-\tme}$ as
\begin{align}
\label{eq:Yop}
\frac{\drm \bm Y_{-\tme}}{\drm \tme}&=-\frac{i}{\hbar}\left[H,\bm Y_{-\tme}\right]+\frac{1}{\hbar^{2}}\left(\mathcal{R}+\mathcal{J}\right),
\end{align}
where
\begin{align}
\mathcal{R}&=\int_{0}^{\tme}\drm \tme_{\sm{1}} [A,\bm Y_{-\tme}A(\tme_{\sm{1}}-\tme)]\ch(\tme_{\sm{1}}-\tme) \nonumber \\
&- [A,A(\tme_{\sm{1}}-\tme)\bm Y_{-\tme}]\ch(\tme-\tme_{\sm{1}}),\\
\mathcal{J}&=i\hbar\int_{0}^{\beta}\drm \beta_{\sm{1}} [A,\bm Y_{-\tme}A(-\tme-i\hbar\beta_{\sm{1}})]\ch(-\tme-i\hbar\beta_{\sm{1}}),
\end{align}
where $\ch(\tme)= \avl E(\tme)E \avr$ and we have taken $\avl E \avr = 0$, which is valid for an environment composed of harmonic oscillators. Above, we have replaced $\tilde{\bm Y}^{{\sm \text{tot}}}(-\tme) \equiv \bm Y^{{\sm \text{tot}}}_{-\tme}$ in the second-order terms since we are interested in the weak-coupling regime. Casting Eq.~(\ref{eq:Yop}) in the energy eigenbasis of the system Hamiltonian $H$, we obtain our central result for the equation of motion as
\begin{align}
\frac{\drm\bm Y_{nm}}{\drm \tme} &=-i\Omega_{nm}Y_{nm}+\frac{1}{\hbar^{2}}\sum_{k,l}\left(\mathcal{R}_{nm}^{kl}+\mathcal{J}_{nm}^{kl}\right)Y_{kl},
\label{eq:master3}\end{align}
where $\hbar \Omega_{nm} = \epsilon_n-\epsilon_m$, with $\epsilon_{n}$'s being the systems' eigenenergies, and
\begin{align}
\mathcal{R}_{nm}^{kl} &= A_{nk}A_{lm}\left[W_{nk}(0,\tme)+W_{ml}^{*}(0,\tme)\right] \nonumber \\
&-\delta_{l,m} \sum_{j} A_{nj}A_{jk}W_{jk}(0,\tme)-\delta_{n,k} \sum_{j} A_{lj} A_{jm}W_{jl}^{*}(0,\tme),\\
\mathcal{J}_{nm}^{kl} &= A_{nk}A_{lm}\left[W_{ml}^{*}(\tme,\infty)-\e^{\beta\hbar\Omega_{lm}}W_{lm}(\tme,\infty)\right] \nonumber \\
&-\delta_{n,k}\sum_{j}A_{lj}A_{jm}\left[W_{jl}^{*}(\tme,\infty)-\e^{\beta\hbar\Omega_{lj}}W_{lj}(\tme,\infty)\right].
\end{align}
We recall that the eigenenergies $\epsilon_n$, as well as the coefficients $\mathcal{R}_{nm}^{kl}$, $\mathcal{J}_{nm}^{kl} $, $W_{nk}(0,\tme)$, all depend parametrically on $\bm B$. The contribution $\mathcal{J}$ in Eq.~(\ref{eq:Yop}) accounts for the salient thermal equilibrium correlations between system and environment. In this initial value term $\mathcal{J}_{nm}^{kl}$ we have converted the imaginary-time integral for the environment correlators to real time using the standard Kubo scheme (see in Ref.~\cite{Hu93AJP61}). Note that the operator $\bm Y_{-\tme}$ above does not obey the basic properties of reduced density operator, i.e., it is not trace preserving [$\Tr_{S}(\bm Y_{-\tme}) \neq 1$], and it need not be positive. Hence, it is worth stressing here that the above EOM is \emph{not} a master equation for the reduced density operator. In order to derive the EOM above, we solely made use of the weak system-environment coupling approximation. The non-Markovian nature of this EOM, however, is evident from the $\tme$ dependence in the $W$ values, i.e., no Markov approximation has been used. 

\begin{figure*}[t!]
\includegraphics[trim= 20mm 0mm 0mm 0mm, width=.7\linewidth]{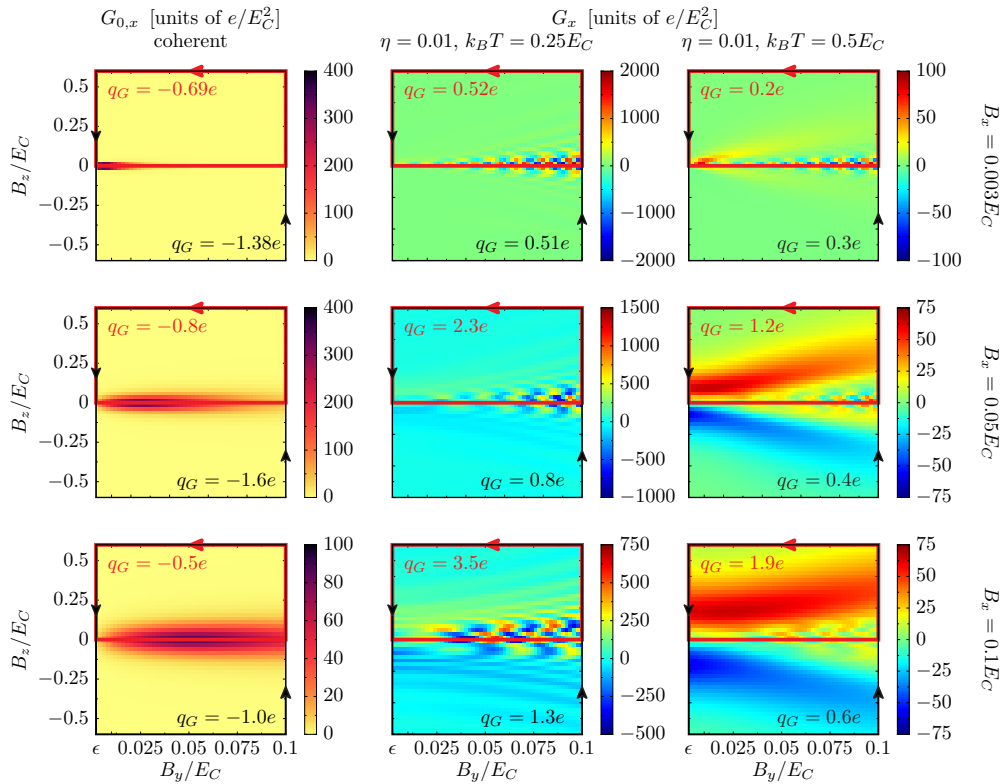}
\caption{(Color online) Geometric  dissipative quantum pumping $q_G$ made visible. The $x$ component of the curl $\bm G$
of the pumping field $\bm F$
on $B_yB_z$ planes at various fixed $B_x$. The plots in the first, second, and third row have
$B_x=0.003 E_C$, $B_x=0.05 E_C$, and $B_x=0.1 E_C$, respectively. The plots in the first column are for the coherent case.
The plots in the second and third
column are for the dissipative case at $\eta=0.01$, and have $k_B T = 0.25 E_C$ and $k_B T=0.5E_C$,
respectively. The quantities
$q_G$, printed in black and red colors, denote the charge pumped on the respective
black and red paths.
The phase $\phi$ is fixed at the value $\phi=\pi/2$. Here $\epsilon=0.002$ sets
the minimal value of $B_y/E_C$. For $E_C\simeq 1 k_B\, \mathrm{K} $ the plots  correspond to
typical experimental ranges.}
\label{fig:fig2}
\end{figure*}

The matrix $W_{ij}(\tme_{\sm{1}},\tme_{\sm{2}})$ characterizes the properties of the environment and can be expressed as $W_{ij}(\tme_{\sm{1}},\tme_{\sm{2}}) = \int_{\tme_{\sm{1}}}^{\tme_{\sm{2}}}\drm\tme \e^{-i\Omega_{ij}\tme}\ch(\tme)$. In order to evaluate the operators $W$ we would require the correlators $\ch(\tme)$, which can be expressed in terms of the spectral density $\sd(\w)$ of the environment as
\begin{align}
\ch(\tme)&=\frac{\hbar}{\pi}\int_{0}^{\infty}\drm\w \sd(\w) \left[\mathrm{coth}\left(\frac{\be\hbar\w}{2}\right)\mathrm{cos}(\w\tme)-\iu \mathrm{sin}(\w\tme)\right].
\end{align}
In case of the ohmic spectral density with Lorentz-Drude cutoff, i.e., $\sd(\w) = \eta \hbar \w\left[1+(\w/\w_{D})^{2}\right]^{-1}$, the correlator can be analytically obtained as
\begin{align}
\ch(\tme) &= \frac{\eta\hbar^{2}}{2}\w_{D}^{2}\left[\mathrm{cot}\left(\frac{\beta\hbar\w_{D}}{2}\right)-i\mathrm{sgn}(\tme)\right]e^{-\w_{D}\tme}\nonumber \\
&-\frac{2\hbar\eta}{\beta}\sum_{l=1}^{\infty}\frac{\nu_{l}}{1-\left(\nu_{l}/\w_{D}\right)^{2}}e^{-\nu_{l}\tme} ~~~~~ \tme \geq 0,
\end{align}
where $\mathrm{sgn}(\tme) = 1$ if $\tme > 0$, or $\mathrm{sgn}(\tme)=0$ if $\tme = 0$ and $\nu_{l} = 2\pi l/(\hbar\beta)$ are the Matsubara frequencies. Using the above form of the correlator, the elements of the $W$ matrix can be readily evaluated, thus forming the relaxation ($\mathcal{R}$) and the initial value ($\mathcal{J}$) tensors. We then propagate the operator $\bm Y$ using a fourth-order Runge-Kutta propagation scheme with special care taken of the initial condition. Initially, the operator $\bm Y_{0} = \Delta \bm S \Tr_{E}\left( \rho^\text{eq}\right)$, where $\Delta \bm S = \bm S - \Tr\left(\rho^\text{eq}\bm S\right)$. Consistent with our weak-coupling approximation, we expand the initial condition up to second order in the coupling strength, using canonical perturbation theory \cite{Thingna12JCP19}.


\section{Results}
Figure \ref{fig:fig2}
presents various density plots of the $x$ component of the curl field on various planes of constant $B_x$.
The first column refers to the coherent regime, Eq. (\ref{eq:G0}), whereas the last two columns refer to the
dissipative regime at two different temperatures, as obtained from solving Eq. (\ref{eq:master3}). For the dissipative regime, gate noise is modeled through $H_{SE}=\sigma_z \otimes \sum c_n x_n$ with coupling
coefficients $c_n$ \cite{Makhlin73RMP01}. For the environment, we chose an ohmic spectral
density $J(\omega)$ with a Lorentz-Drude cutoff $\omega_D$:  $J(\omega)={\eta \hbar \omega}[1+
(\omega/\omega_D)^2]^{-1}$. Here $\eta$ determines the dissipation strength.
Printed in black (red) are the values, $q_G$, of the geometrically pumped charge on a path
encircling the whole graph (black paths) and half graph (red paths), respectively.

\begin{figure*}[t!]
		\includegraphics[width=0.7\linewidth]{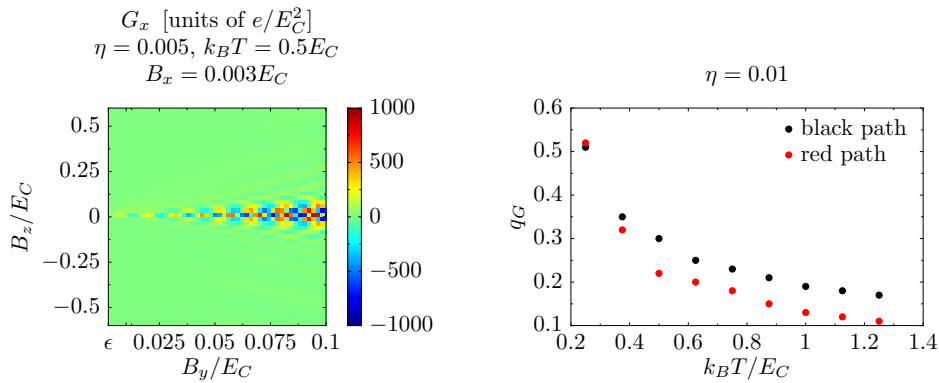}
	\caption{(Color online) The left panel shows the $x$ component of the curl of the pumping field $\bm F$ on the $B_{y}B_{z}$ plane for a fixed $B_{x} = 0.003 E_{C}$ at $k_{B}T = 0.5E_{C}$ and $\eta = 0.005$. The right panel shows the pumped charge $q_{G}$ for the red path (red dots) and the black path (black dots) (see Fig.~\ref{fig:fig2}) for $B_{x} = 0.003 E_{C}$ and $\eta = 0.01$. All other parameters are the same as in Fig.~\ref{fig:fig2}.}
     \label{fig:3}
\end{figure*}

Striking differences emerge between the two regimes. The most apparent is the emergence of an asymmetry of $G_x$ under $B_z \leftrightarrow -B_z$, in the dissipative regime as opposed to the coherent regime. This is due to the fact that the two charge states
$|0\rangle$ and  $|1\rangle$ are differently coupled to the bath. 
At finite dissipation the asymmetry is
weak for small values of $B_z$, while at large $B_z$ values, it even turns (at least approximately) into an
odd symmetry. Interestingly, this can be used to enhance the
pumped charge. Take, for example, paths that enclose half of the graph, cf. the red paths in Fig.
\ref{fig:fig2}. At finite dissipation, they pump more than the paths enclosing the whole graph, where
upper and lower parts contribute with opposite signs to the pumped charge. One can pump as
much as $3.5\, e$ per cycle on the red path at $B_x=0.1 E_C$, $T= 0.25 E_C$. The same path
would pump as little as $-0.5\,  e$ in the zero dissipation case. This evidences the beneficial role of
dissipation in quantum pumping. Dissipation can even give rise  to a change of direction of the current. To
conceive this  current reversal we shall recall that the force response $\langle \Delta \bm
S_t\rangle$ is composed of two terms: the friction $\bm \gamma$
and a geometric magnetism $\bm{\mathcal B}$
\cite{Berry93PRSLA442,Nulton85JCP83,Servantie03PRL91,Koning05JCP122,Sivak12PRL108,Bonanca14JCP140,Campisi12PRA86}:
 $\langle \Delta \bm S_t\rangle = - \bm \gamma \cdot
\dot{\bm B} -\bm{\mathcal B} \times \dot{\bm B} $, respectively given by the symmetric and
antisymmetric component of the conductance matrix, $K_{ij}=-\int_{-\infty}^0 ds \int_0^\beta du
\langle  S^i_{-i\hbar u} \Delta  S_s^j \rangle_{\bm B}^\text{eq}$, according to the formulas $
\gamma_{ij}= K_{ij}^S$, $\mathcal B_{k}=-\sum_{ij}\varepsilon_{ijk}K_{ij}^A/2$ [where $\varepsilon_{ijk}$
is the Levi-Civita symbol and $K_{ij}^{S(A)}=(K_{ij}\pm K_{ji})/2$]. We recall that $\bm \gamma$ is exactly null in
the coherent regime. Noticing that the current
operator, Eq. (\ref{eq:I}), is a linear combination of $S_x$ and $S_y$ allows us to express the pumping
field $\bm F$ as a linear combination of the $K_{ij}$ values or, accordingly, of their symmetric and
antisymmetric parts, $K_{ij}^S$ and $K_{ij}^A$. This in turn allows us to quantify the fractions of
pumped charge due to geometric magnetism and friction, respectively. We have found that both
contributions are greatly affected by the presence of the thermal bath but appear to have
competing roles, i.e., they possess opposite signs. On the red paths in Fig. \ref{fig:fig2}, friction wins over
geometric magnetism at finite $\eta$, thus resulting in a different current direction as compared to
the coherent case where only geometric magnetism is present \cite{Berry93PRSLA442}.

Each pixel in the graphs presented in Fig. \ref{fig:fig2} is the result of a single simulation at the corresponding value $\bm B$ of the parameters. In Fig.~\ref{fig:3} (left panel) we report additional results for one particular slice of $k_{B} T = 0.5 E_{C}$, and $B_{x} = 0.003 E_{C}$ for a weaker system-environment coupling strength, $\eta=0.005$, as compared to the coupling strength $\eta = 0.01$ in Fig. \ref{fig:fig2}. The right panel of Fig.~\ref{fig:3} depicts the pumped charge $q_{G}$ for $B_{x}=0.003 E_{C}$ and $\eta = 0.01$ for the red and black paths, as shown in Fig.~\ref{fig:fig2}. As expected, the pumped charge decreases with the increase in temperature.

\section{Conclusions}
We have presented a geometric method for calculation of the geometrically pumped charge $q_G$ in
dissipative quantum systems. The method is based on the calculation of the pumping field $\bm F$ and uses
the salient observation that the latter coincides with the vector of linear response coefficients of the adiabatic
expansion of the current [Eq. (\ref{eq:Taylor})]. For a dissipative open quantum system, this is given by equilibrium
quantum correlation functions calculated at fixed driving parameters [Eq. (\ref{eq:F})]. Hence, in contrast to the
customary procedure, they can be conveniently evaluated by solving an undriven problem. Our method for the calculation of $\bm F$ consists in deriving an equation of motion for a properly chosen observable under only the assumption of weak coupling (no Markov approximation, no rotating wave approximation, no factorized initial condition). The timely application to the Cooper-pair sluice reveals an interplay of geometric magnetism
and a dissipation-induced enhancement of pumped charge, the emergence of current reversals, and asymmetries. All these novel phenomena can {\it a priori} be identified visually upon the
mere inspection of the pumping field in control parameter space.
Most importantly, the results presented in our  Fig. \ref{fig:fig2} can be experimentally checked with current devices and setups.

\section*{Acknowledgments}
The authors thank Jukka Pekola for useful remarks.
This research was supported by a Marie Curie Intra European Fellowship within the 7th European Community Framework
Programme through the projects NeQuFlux, Grant No. 623085  (M.C.), and ThermiQ, Grant No. 618074 (R.F.), by the COST action MP1209
through a Short Term Scientific Mission (M.C.); by MIUR-PRIN, ``Collective quantum phenomena: From strongly correlated
systems to quantum simulators'' (R.F.), and by the Volkswagen Foundation, Project No. I/83902 (P.H., M.C.).
This work is dedicated to the memory of late Prof. Donald H. Kobe.


\begin{thebibliography}{53}

\bibitem{Berry84PRSLA392}
M.V. Berry, Proc. R. Soc. London, Ser. A \textbf{392}, 45 (1984).

\bibitem{Resta94RMP66}
R.~Resta, Rev. Mod. Phys. \textbf{66}, 899 (1994).

\bibitem{Xiao10RMP10}
D.~Xiao, M.C. Chang, Q.~Niu, Rev. Mod. Phys. \textbf{82}, 1959 (2010).

\bibitem{Sinitsyn09JPA42}
N.A. Sinitsyn, J. Phys. A: Math. Theo. \textbf{42}, 193001 (2009).

\bibitem{Wilczek89Book}
\emph{Geometric Phases in Physics}, edited by F.~Wilczek, A.~Shapere, Advanced Series in Mathematical Physics Vol. 5 (World Scientific, Singapore, 1989).

\bibitem{Hannay06AJP74}
J.H. Hannay, Am. J. Phys. \textbf{74}, 134 (2006).

\bibitem{Hanggi09RMP81}
P.~H{\"a}nggi, F.~Marchesoni, Rev. Mod. Phys. \textbf{81}, 387 (2009).

\bibitem{Ren10PRL104}
J.~Ren, P.~H{\"a}nggi, B.~Li, Phys. Rev. Lett. \textbf{104}, 170601 (2010).

\bibitem{Thouless83PRB27}
D.J. Thouless, Phys. Rev. B \textbf{27}, 6083 (1983).

\bibitem{Brouwer98PRB58}
P.W. Brouwer, Phys. Rev. B \textbf{58}, R10135 (1998).

\bibitem{Zhou99PRL82}
F.~Zhou, B.~Spivak, B.~Altshuler, Phys. Rev. Lett. \textbf{82}, 608 (1999).

\bibitem{Pekola13RMP85}
J.P. Pekola, O.P. Saira, V.F. Maisi, A.~Kemppinen, M.~M\"ott\"onen, Y.A.
  Pashkin, D.V. Averin, Rev. Mod. Phys. \textbf{85}, 1421 (2013).

\bibitem{Aleiner98PRL81}
I.L. Aleiner, A.V. Andreev, Phys. Rev. Lett. \textbf{81}, 1286 (1998).

\bibitem{Entin-Wohlman02PRB65}
O.~Entin-Wohlman, A.~Aharony, Y.~Levinson, Phys. Rev. B \textbf{65}, 195411
  (2002).

\bibitem{Moskalets02PRB66}
M.~Moskalets, M.~B\"uttiker, Phys. Rev. B \textbf{66}, 205320 (2002).

\bibitem{Citro03PRB68}
R.~Citro, N.~Andrei, Q.~Niu, Phys. Rev. B \textbf{68}, 165312 (2003).

\bibitem{Splettstoesser05PRL95}
J.~Splettstoesser, M.~Governale, J.~K\"onig, R.~Fazio, Phys. Rev. Lett.
  \textbf{95}, 246803 (2005).

\bibitem{Sela06PRL96}
E.~Sela, Y.~Oreg, Phys. Rev. Lett. \textbf{96}, 166802 (2006).

\bibitem{Fioretto08PRL100}
D.~Fioretto, A.~Silva, Phys. Rev. Lett. \textbf{100}, 236803 (2008).

\bibitem{Pekola01PRL64}
J.P. Pekola, J.J. Toppari, Phys. Rev. B \textbf{64}, 172509 (2001).

\bibitem{Governale05PRL95}
M.~Governale, F.~Taddei, R.~Fazio, F.W.J. Hekking, Phys. Rev. Lett.
  \textbf{95}, 256801 (2005).

\bibitem{Mottotten06PRB73}
M.~M\"ott\"onen, J.P. Pekola, J.J. Vartiainen, V.~Brosco, F.W.J. Hekking, Phys.
  Rev. B \textbf{73}, 214523 (2006).

\bibitem{Onoda06PRL97}
S.~Onoda, C.H. Chern, S.~Murakami, Y.~Ogimoto, N.~Nagaosa, Phys. Rev. Lett.
  \textbf{97}, 266807 (2006).

\bibitem{Pothier92EPL17}
H.~Pothier, P.~Lafarge, C.~Urbina, D.~Esteve, M.H. Devoret, Europhys. Lett. \textbf{17}, 249 (1992).

\bibitem{Mottonen08PRL100}
M.~M\"ott\"onen, J.J. Vartiainen, J.P. Pekola, Phys. Rev. Lett. \textbf{100},
  177201 (2008).

\bibitem{Pekola08NATPHYS4}
J.P. Pekola, J.J. Vartiainen, M.~Mottonen, O.P. Saira, M.~Meschke, D.V. Averin,
  Nat. Phys. \textbf{4}, 120 (2008).

\bibitem{Giazotto11NATPHYS7}
F.~Giazotto, P.~Spathis, S.~Roddaro, S.~Biswas, F.~Taddei, M.~Governale,
  L.~Sorba, Nat. Phys. \textbf{7}, 857 (2011).

\bibitem{Grifoni98PR304}
M.~Grifoni, P.~H\"anggi, Phys. Rep. \textbf{304}, 229 (1998).

\bibitem{Brandes02PRB66}
T.~Brandes, T.~Vorrath, Phys. Rev. B \textbf{66}, 075341 (2002).

\bibitem{Salmilehto10PRA82}
J.~Salmilehto, P.~Solinas, J.~Ankerhold, M.~M\"ott\"onen, Phys. Rev. A
  \textbf{82}, 062112 (2010).

\bibitem{Solinas10PRB82}
P.~Solinas, M.~M\"ott\"onen, J.~Salmilehto, J.P. Pekola, Phys. Rev. B
  \textbf{82}, 134517 (2010).

\bibitem{Pekola10PRL105}
J.P. Pekola, V.~Brosco, M.~M\"ott\"onen, P.~Solinas, A.~Shnirman, Phys. Rev.
  Lett. \textbf{105}, 030401 (2010).

\bibitem{Russomanno11PRB83}
A.~Russomanno, S.~Pugnetti, V.~Brosco, R.~Fazio, Phys. Rev. B \textbf{83},
  214508 (2011).

\bibitem{Kamleitner11PRB84}
I.~Kamleitner, A.~Shnirman, Phys. Rev. B \textbf{84}, 235140 (2011).

\bibitem{Pellegrini11PRL107}
F.~Pellegrini, C.~Negri, F.~Pistolesi, N.~Manini, G.E. Santoro, E.~Tosatti,
  Phys. Rev. Lett. \textbf{107}, 060401 (2011).

\bibitem{Wollfarth13PRB87}
P.~Wollfarth, I.~Kamleitner, A.~Shnirman, Phys. Rev. B \textbf{87}, 064511
  (2013).

\bibitem{footnote}
For example, a fluidic current flows through a peristaltic pump if none of the pistons fully obstructs the channel in presence of a  pressure gradient. Similarly a superconducting current flows through a Cooper pair sluice in presence of a phase bias.

\bibitem{Aunola03PRB68}
M.~Aunola, J.J. Toppari, Phys. Rev. B \textbf{68}, 020502 (2003).

\bibitem{Berry93PRSLA442}
M.V. Berry, J.M. Robbins, Proc. R. Soc. London, Ser. A \textbf{{442}}, 659
  ({1993}).

\bibitem{Campisi12PRA86}
M.~Campisi, S.~Denisov, P.~H\"anggi, Phys. Rev. A \textbf{86}, 032114 (2012).

\bibitem{BreuerPetruccioneBOOK}
H.P. Breuer, F.~Petruccione, \emph{The Theory of Open Quantum Systems} (Oxford
  University Press, Oxford, UK, 2002).

\bibitem{Talkner86AP167}
P.~{Talkner}, Ann. Phys. \textbf{167}, 390 (1986).

\bibitem{Caldeira83AP149}
A.O. {Caldeira}, A.J. {Leggett}, Ann. Phys. (N.Y.) \textbf{149}, 374 (1983).

\bibitem{Hanggi05CHAOS15}
P.~H{\"a}nggi, G.~Ingold, Chaos \textbf{{15}}, 026105 ({2005}).

\bibitem{Makhlin73RMP01}
Y.~Makhlin, G.~Sch\"on, A.~Shnirman, Rev. Mod. Phys. \textbf{73}, 357 (2001).

\bibitem{Hu93AJP61}
B.Y.K. Hu, Am. J. Phys. \textbf{61}, 457 (1993).

\bibitem{Grabert80JSP22}
H.~Grabert, P.~H\"anggi, P.~Talkner, J. Stat. Phys. \textbf{22}, 537 (1980).

\bibitem{Talkner09JSM09}
P.~Talkner, M.~Campisi, P.~H{\"a}nggi, J. Stat. Mech. (2009) P02025.

\bibitem{Thingna12JCP19}
J.~Thingna, J.S. Wang, P.~H{\"a}nggi, J. Chem. Phys. \textbf{136}, 194110
  (2012).

\bibitem{Nulton85JCP83}
J.~Nulton, P.~Salamon, B.~Andresen, Q.~Anmin, J. Chem. Phys. \textbf{83}, 334
  (1985).

\bibitem{Servantie03PRL91}
J.~Servantie, P.~Gaspard, Phys. Rev. Lett. \textbf{91}, 185503 (2003).

\bibitem{Koning05JCP122}
M.~de~Koning, J. Chem. Phys. \textbf{122}, 104106 (2005).

\bibitem{Sivak12PRL108}
D.A. Sivak, G.E. Crooks, Phys. Rev. Lett. \textbf{108}, 190602 (2012).

\bibitem{Bonanca14JCP140}
M.V.S. Bonanca, S.~Deffner, J. Chem. Phys. \textbf{140}, 244119 (2014).

\end{thebibliography}
\end{document}